\documentstyle[prl,aps,12pt]{revtex}
\input{epsf.sty}

\def\b\mu{{\bf \mu}}

\begin{document}

\draft

\title{\Large \bf Symmetry Nonrestoration in a Resummed Renormalized 
Theory at High Temperature}
\vspace{1cm}
\author{H. C. de Godoy Caldas\thanks{e-mail address: hcaldas@funrei.br}}

\address{Departamento de Ci\^{e}ncias Naturais, DCNAT \\
Universidade Federal de S\~{a}o Jo\~{a}o del Rei, UFSJ \\ 
Pra\c{c}a Dom Helv\'ecio, 74, CEP:36300-000, S\~{a}o Jo\~{a}o del Rei, MG, Brazil} 

\date{May, 2002}

\maketitle                          

\vspace{0.5cm}

\begin{abstract}
We reinvestigate the interesting phenomenon of symmetry nonrestoration at high temperature in the multifield $O(N_1) X O(N_2)$ model. We apply modified self-consistent resummation (MSCR) in order to obtain the scalar dressed masses and find in what circumstances a resummed multifield theory which has symmetry (non)restoration can be renormalized. It is shown that, aside from the consistency of the MSCR method, the basic ingredient that guarantees the renormalization of a multifield model within a resummation approach is the $T^2$ mass behavior of field theory at high temperature.
\end{abstract}

\vspace{2cm}


\newpage

\section{Introduction} 

It was first noticed by Linde in Ref. \cite{Lin} that broken symmetries may be restored at high temperature. Indeed, at very high temperatures, the leading-order corrections are enough to restore the broken symmetries in most of the conventional models \cite{Dolan,Weinberg}. However, Weinberg has shown that, in some models such as the $O(N_1)XO(N_2)$ scalar theory, the symmetry is not restored as the temperature increases when the negative coupling between the fields extrapolates some limiting value \cite{Weinberg}. This intriguing phenomenon is called symmetry nonrestoration (SNR) [or inverse symmetry breaking (ISB)] and has not only academic interest. Symmetry nonrestoration at high temperature has crucial consequences in the cosmological scenario and could, in principle, avoid the monopole problem \cite{Goran1,Bimonte1}. Some examples of SNR in nature and experimental discoveries may be found, e.g., in \cite{Bajac}. Nevertheless, even in theories where SNR is believed to be found, there has been some controversy if the phenomenon, in fact, happens as it was formulated. See, for instance, Ref. \cite{Marcus1} for a discussion. Recent works show that higher order terms or a nonperturbative study tend to diminish the parameter space where symmetry nonrestoration can happen \cite{Marcus1,Bimonte}. Reference \cite{Bimonte} is a one loop self consistent resummation calculation whereas Ref. \cite{Marcus1} takes full account of two loop diagrams in a nonperturbative fashion. The parameter space for ISR predicted by the latter is bigger than the one predicted by the former. Also the two loop calculation reveals a new possibility for ISB where the last term in the breaking sequence is $O(N_1-1) X O(N_2-1)$ which cannot be seen at the one loop level \cite{Marcus1}.

In this paper we will apply a nonperturbative method, modified self-consistent resummation (MSCR), which we have developed recently \cite{Caldas1,Caldas2,Caldas3} in order to reinvestigate some aspects of SNR. To motivate, let us note that one way to justify the study of resummation programs in systems which exhibit symmetry (non)restoration is that those methods represent a practical way to recover the reliability of perturbation theory at finite temperatures. However, another important reason is that these methods can be a natural manner to take into account the medium-modified masses in the computation of several physical processes such as dispersion relations, damping rates and decay widths. The goal of MSCR is to make renormalization possible when obtaining the dressed masses in a nonperturbative fashion. We will follow a different direction from the conventional approaches about symmetry nonrestoration at high temperature. These are, in general, concerned mainly with the behavior of the modulus of the coupling between the fields in the region of symmetry nonrestoration. Various nonperturbative methods have been used to study symmetry nonrestoration, but the important issue of renormalization in these kinds of multifield theories has not been discussed, as in Ref. \cite{Bimonte}. Although very important, the details concerning the implementation of an appropriate renormalization program within those methods have been overlooked in most applications. We focus on the necessary conditions for a resummed multifield theory that has its symmetry restored or not at high temperature, to be satisfactorily renormalized.

To assure renormalizability in a finite temperature application of resummation programs to theories which have fields with different masses, one must require that the masses obey two essential features.\\ 

{\bf (1)} A combination that relates them (which can be imposed by some symmetry as in the linear sigma model where $M_{\sigma}^2 = M_{\pi}^2 + 2\lambda\nu^2$) \cite{Caldas1}. However, as we shall see soon, this does not necessarily mean that the symmetry is restored at high temperature.\\ 

{\bf (2)} The masses in the internal lines of the diagrams have to be the same as in the counterterms. These requirements are because the resummation solves the infrared divergences problem (or the problem of higher order terms in higher order calculations) but may bring another one: the divergent temperature dependent parts of the diagrams do not match the same terms in the original Lagrangian. That is, the usual constant counterterms cannot eliminate terms which are functions of the temperature. At zero temperature \cite{Franco} or in the naive one-loop calculations at finite temperature \cite{Caldas4} this point does not represent a problem since the masses running in the loops are also constants. Then, in resummation methods as MSCR it is crucial that the Lagrangian, including the counterterms, suffer the effects of the resummation for the model to be satisfactorily renormalizable \cite{Caldas1,Caldas3,Parwani}. The second necessary condition is naturally fulfilled by MSCR since the method is based on the recalculation of the self-energy using in each step the masses obtained in the previous one. In this way, the absorption of the unwanted ultraviolet divergences is always guaranteed once the masses running in the loops are necessarily the same as in the counterterms. Regarding the first condition, in Sec. III we show that we can construct a renormalizable theory (in the sense we are discussing) that exhibits both symmetry restoration and nonrestoration, depending on the values given to the parameters which govern the relation between the masses.

This paper is organized as follows. The application of the MSCR method in the $O(N_1) X O(N_2)$ model is addressed in Sec. II. In Sec. III we calculate the dressed masses and determine the necessary conditions for a resummed multifield theory which displays symmetry restoration or nonrestoration 
at high temperature to be renormalized. In Sec. IV we discuss some numerical results. Section V is devoted to the conclusions.

\section{The $O(N_1) X O(N_2)$ Model}

Consider the $O(N_1) X O(N_2)$ model described by the Lagrangian

\begin{equation}
\label{eq1}
{\cal L}={\cal L}^{0}(\phi)+ {\cal L}^{int}(\phi) + {\cal L}^{ct}(\phi),
\end{equation}
where
\begin{equation}
\label{eq2}
{\cal L}^0={1\over2}\sum_{i=1}^2 \left[(\partial_\mu\phi_i)^2 -
 m_i^2\phi_i^2 \right],
\end{equation}
the interaction Lagrangian is expressed as 
\begin{equation}
\label{eq3}
{\cal L}^{int}= - \sum_{i=1}^2 \left[ {\lambda_i \over 4!}\phi_i^4 \right]
- {\lambda \over 4}\phi_1^2 \phi_2^2,
\end{equation}
and the counterterm Lagrangian which is necessary to render the theory finite up to a given order is expressed as

\begin{equation}
\label{eq4}
{\cal L}^{ct}= \sum_{i=1}^2 \left[{1\over2} A (\partial_\mu\phi_i)^2 -{1\over2} B_i m_i^2 \phi_i^2  
- {1\over 4!} C_i \phi_i^4 \right] 
- {1\over 4} C \phi_1^2 \phi_2^2. 
\end{equation}

The conditions required for the model described by Eq. (\ref{eq1}) to be bounded from below are 
\begin{equation}
\label{eq4.1}
\lambda_1 > 0,~~~~\lambda_2 >0~~~~ {\rm and}~~~~ \lambda_1 \lambda_2 > 9 \lambda^2.
\end{equation}

Let us now apply some topics of MSCR in the $O(N_1) X O(N_2)$ model. The MSCR recipe \cite{Caldas1,Caldas3} dictates that 

\begin{equation}
\label{eq4.2}
M_{i,n} ^2= (B_{i,n}+1)M_{i,n-1} ^2 + \Pi_i(K_0,|\vec{K}|=0,M_{i,n-1}),
\end{equation}
where $n$ is the order of the nonperturbative correction, $B_n$ is the coefficient of the appropriate counterterm and $\Pi_i(K_0,|\vec{K}|=0,M_{i,n-1})$ is the self-energy of the field $i$ in a given order in the perturbative expansion at zero external three momentum. This assures the cancellation of the ultraviolet divergences since the masses running in loops are necessarily the same as in the counterterms. The worse divergences i.e., the infrared, will be cutoff by the recalculation (resummation) of the self-energy, which is achieved for $n >1$. Then, the nonperturbative correction to the thermal mass of the field $\phi_i~(i=1,2)$ to one-loop order, which is independent of the external momentum, is given by

\begin{eqnarray}
\label{eq5}
M_{1,n}^2 (T)= (B_{1,n}+1)M_{1,n-1} ^2 -\frac{\lambda_1}{2(4\pi)^2} \left(\frac{N_1+2}{3} \right) M_{1,n-1}^2 (T) \frac{1}{\tilde \epsilon} \\
\nonumber
-\frac{\lambda}{2(4\pi)^2} N_2~ M_{2,n-1}^2 (T) \frac{1}{\tilde \epsilon}
+ Finite_1 + \Pi_1^{\beta}(M_{1,n-1}^2 (T),M_{2,n-1}^2 (T)),
\end{eqnarray}

\begin{eqnarray}
\label{eq6}
M_{2,n}^2 (T)= (B_{2,n}+1)M_{2,n-1} ^2 -\frac{\lambda_2}{2(4\pi)^2} \left(\frac{N_2+2}{3} \right) M_{2,n-1}^2 (T) \frac{1}{\tilde \epsilon} \\
\nonumber
-\frac{\lambda}{2(4\pi)^2} N_1~ M_{1,n-1}^2 (T) \frac{1}{\tilde \epsilon} +
Finite_2 + \Pi_2^{\beta}(M_{1,n-1}^2 (T),M_{2,n-1}^2 (T)),
\end{eqnarray}
where $ Finite_i$ is the finite part of the divergent contributions, $\Pi_i^{\beta}$ is the temperature-dependent contribution and $\frac{1}{\widetilde \epsilon} \equiv \frac{2}{4-d}-\gamma + \log(4\pi)$ (this is the modified minimal subtraction scheme used in dimensional regularization that we have adopted here), where $\gamma$ is the Euler constant.

Let us stop the calculations in the $O(N_1) X O(N_2)$ theory for a while 
and turn the attention to a more simple case, in order to compare some results of the MSCR method with another approach \cite{Tanguy}. In Ref. \cite{Tanguy}, Altherr computed the self-energy up to two-loops in the $\lambda \phi ^4$ model, and after had identified the origin of the leading infrared part (which came from the double scope diagram), he proposed a possible cure for this infrared singularity. He suggested as a cure the thermal mass obtained in the massless limit of the one-loop order, which could be used as a cutoff. This cutoff mass was put only in the infrared part of the self-energy cited above. He saw that this (part of the two-loop order in the perturbative expansion) manipulated contribution could be gotten by N-loop diagrams. So he gave up the perturbative expansion and considered the most infrared singular diagrams, which are the daisy types, as a nonperturbative contribution to the mass shift. He found that the result of the summation of these N-loop diagrams is not very different from the perturbative approach.

In Ref. \cite{Caldas3}, we also treat the $\lambda \phi ^4$ model up to two loops, but using MSCR. As explained before, this method is based on consistently obtaining the pole of the thermal corrected propagator, up to a given number of loops in the perturbative expansion, totally free from UV 
and IR divergences, due to the resummation nature of the method. This is completely different from what Altherr has done, since with MSCR one does not have to replace the perturbative expansion by a nonperturbative approach. Besides, a nonperturbative approximation has to be done exactly to recover the validity of the perturbation series. As can be seen in Ref. \cite{Caldas3}, the corrections to the free propagator are the one-and two-loop graphs from the perturbative expansion plus infinitely many others obtained nonperturbatively, like superdaisy types. MSCR has the advantage of keeping the same fundamental theory when calculating the pole of the corrected propagator, using the same counterterms as at zero temperature. The only difference is that the mass running in the loops changes in each recalculation (each summation of an infinity set of diagrams) of the self-energy. MSCR is thus more natural and keeps the loop expansion, as in Ref. \cite{Parwani}. With the MSCR, the ``starting'' effective Lagrangian can be chosen such that the mass running in the loops is the one-loop thermal mass (or, in other words, the plasmon mass $m^2=\frac{1}{24}\lambda T^2$), avoiding in this way infrared problems and respecting the symmetries of the Lagrangian, since a loop expansion is an expansion in powers of the Lagrangian. 

Then, MSCR can be understood as an alternative and consistent resummation method, whose renormalization is in the same way as in the zero temperature case. More simple and less complicated naive approaches have found several difficulties when trying to find appropriate renormalization conditions for the divergent gap equations, as discussed in \cite{Caldas1}.

\section{The Dressed Masses in the $O(N_1) X O(N_2)$ Model}

We shall now compute the dressed masses and investigate the basic conditions in order for the multifield (``multimasses'') model to be renormalizable, i.e., completely free from ``temperature dependent'' ultraviolet divergences. To be more clear, $(B_{i,n}+1)M_{i,n-1} ^2$ in Eqs. (\ref{eq5}) and (\ref{eq6}) should be written as $(B_{i,n}+1)m_i^2 + (\tilde B_{i,n}+1)\Pi_i^{Ren}(M_{i,n-2}^2)$, since in these equations $M_{i,n-1} ^2=m_i^2+\Pi_i^{Ren}(M_{i,n-2}^2)$. See the appendix for details. The coefficients of the temperature dependent mass counterterms $\tilde B_{i,n}$ are fixed in a manner to cancel not only divergences proportional to $\Pi_i^{Ren}(M_{i,n-2}^2)$ but also these terms together. This is to avoid overcounting of diagrams\cite{Caldas1}. In order for these temperature-dependent infinities to be canceled properly by ${\cal L}^{ct}$ let us consider the most general case of (linear) combination between the squared masses. Namely, $M_{1,n-1}^2 (T) = aM_{2,n-1}^2 (T)+{\cal F}$, where $a$ may be a constant and ${\cal F}$ is some well behaved function which can, in principle, depend on $(\lambda_i,\lambda,N_1,N_2,T)$ \footnote{Remembering again the linear sigma model, $M_{\sigma}^2 = M_{\pi}^2 + {\cal F}$, with ${\cal F}=2\lambda\nu^2$ ($\nu$ is the VEV of the sigma field), where this relation holds at tree level and in the low temperature region. At high temperature ${\cal F}=0$, signalizing that the symmetry is restored. In the region of intermediate temperatures, i.e., around $T_c$ (the critical temperature) the one-loop analysis is not enough to describe the behavior of the system and the validity of this relation \cite{Caldas1}.} and may be related with the vacuum expectation value (VEV) of a scalar field. For simplicity we will take ${\cal F}$ as being zero and, as we are going to see below, a particular choice for $a$ is $a=\pm 1$ with $a=1$ representing symmetry restoration and $a=-1$, on the contrary, representing symmetry non-restoration. In fact, we will prove in the next section that, at sufficiently high temperature and for realistic values of $N_1$ and $N_2$ and appropriate values the couplings respecting the boundness condition, these exact values for $a$ can be found. We argue that in ordinary field theory this relation is always satisfied, since at high temperature the fields have a $T^2$ mass behavior. To the best of our knowledge, the exception is left to a supersymmetric theory with flat directions which do not have interactions strong enough to be in thermal equilibrium at high temperature. A direct consequence of this is the absence of $T^2$ mass terms\cite{Bajc}.

Since $n \to \infty$ (which makes sense within a resummation program) then $n \simeq n-1$. This implies that 
$M_{1,n}^2 (T) = \pm M_{2,n}^2 (T) = \pm M^2(T)$. After this, the coefficients of the mass counterterms at one-loop order are easily found to be 

\begin{eqnarray}
\label{eq6-a}
B_{1,n}\pm= \frac{\lambda_1}{2(4\pi)^2} \left(\frac{N_1+2}{3} \right) 
\frac{1}{\tilde \epsilon} \pm \frac{\lambda}{2(4\pi)^2} N_2~ 
\frac{1}{\tilde \epsilon} \\
\nonumber 
B_{2,n}\pm= \frac{\lambda_2}{2(4\pi)^2} \left(\frac{N_2+2}{3} \right) 
\frac{1}{\tilde \epsilon} \pm \frac{\lambda}{2(4\pi)^2} N_1~ 
\frac{1}{\tilde \epsilon}.
\end{eqnarray}

Let us now see the consequences of this ``prescription'' on the masses resummed by an infinity set of ``daisy'' and ``superdaisy'' diagrams which are achieved in the $n \to \infty$ limit \cite{Caldas1,Caldas3}. The study will be divided into two cases.\\
\\
\begin{center}
{\bf A. First case: Symmetry restoration ($a=1$)}\\
\end{center}

This is the common case where the symmetry is restored at high temperature and will simply furnish a relation between the couplings

\begin{eqnarray}
\label{eq7}
m_1^2 + \left[ \lambda_1 \left( \frac{N_1 + 2}{3} \right) + \lambda N_2 \right]\frac{T^2}{24} 
\left(1- \frac{3M}{\pi T} \right)=\\
\nonumber 
m_2^2 + \left[ \lambda_2 \left( \frac{N_2 + 2}{3} \right) + \lambda N_1 \right]\frac{T^2}{24} 
\left(1- \frac{3M}{\pi T} \right),
\end{eqnarray}
which at high temperature gives for both positive and negative $\lambda$,

\begin{equation}
\label{eq8}
\lambda_1= \left( \frac{N_2 + 2}{ N_1 + 2} \right) \lambda_2 \pm 
3\left( \frac{N_1 - N_2}{ N_1 + 2} \right)|\lambda|.
\end{equation}

Equation (\ref{eq8}) shows the result that for $N_1=N_2$ we have $\lambda_1=\lambda_2$. Although $\lambda$ here can assume negative values, it is not completely free to be larger than the coefficient of $T^2$ in Eq. (\ref{eq7}). This would imply a nonphysical theory since the third condition imposed for the model to be bounded from below in Eq. (\ref{eq4.1}) would not be satisfied, i.e., one would have $\lambda_1\lambda_2<9\lambda^2\frac{N_1N_2}{(N_1+2)(N_2+2)}$.\\
\\
\begin{center} 
{\bf B. Second case: Symmetry nonrestoration ($a=-1$)}\\
\end{center}

In this situation we have $M_1^2=-M_2^2$. One can choose $M_2^2$ as being the mass which has the negative coefficient such that

\begin{equation}
\label{eq9}
\lambda_2 \frac{N_2 + 2}{3} + \lambda N_1 < 0,
\end{equation}
so, inverse symmetry breaking will happen for 

\begin{equation}
\label{eq10}
|\lambda| > \frac{\lambda_2}{N_1}\left( \frac{N_2+2}{3} \right). 
\end{equation}

As we will see below, the exact constraints $a=1$ as well as $a= - 1$ can be found for realistic values of the parameters of the model, respecting the boundness condition expressed by the inequalities (\ref{eq4.1}).

\section{Numerical Analysis}
\label{Num} 

With the intention of verifying that the approximations imposed on the renormalized theory indeed describe very well the cases of symmetry restoration and nonrestoration at high temperature let us define new variables as

\begin{equation}
\label{eq11}
\tilde M_1^2 \equiv M_1^2 - m_1^2 = \left[ \lambda_1 \left( \frac{N_1 + 2}{3} \right) + \lambda N_2 \right]\frac{T^2}{24}.
\end{equation}

\begin{equation}
\label{eq12}
\tilde M_2^2 \equiv M_2^2 - m_2^2 = \left[ \lambda_2 \left(\frac{N_2 + 2}{3} \right) + \lambda N_1 \right]\frac{T^2}{24},
\end{equation}

The first case, $\tilde M_1^2=\tilde M_2^2$, that is related to usual symmetry restoration at high temperature takes place if one chooses, for example, $N_1=90$, $N_2=24$ (where we have followed Ref. \cite{Bimonte} by choosing these values in order the model mimics the Higgs sector of a $SU(5)$ grand unified model), $\lambda_2=0.83$ and $\lambda=+0.09$ which after put in Eq. (\ref{eq8}) gives $\lambda_1=0.43$ and the exact value $a=1$, as expected.

To achieve the second case, we have chosen $\lambda_2\left( \frac{N_2 + 2}{3} \right) - |\lambda| N_1 < 0$, so that $\tilde M_2^2$ is negative. We also chose $N_1=90$ and $N_2=24$ and $\lambda_1=0.1$, $\lambda_2=0.83$ and $\lambda=-0.09$ such that the conditions (\ref{eq4.1}) and (\ref{eq10}) are still satisfied even with the cross coupling $\lambda$ negative. With these values chosen, we obtain exactly the wanted result that describes symmetry nonrestoration $a= \frac{\left[ \lambda_1 \frac{N_1 + 2}{3} - |\lambda| N_2 \right]}{\left[ \lambda_2 \frac{N_2 + 2}{3} - |\lambda| N_1 \right]}=-1$ and consequently, $\tilde M_1^2=-\tilde M_2^2$. In Fig. 1 one sees the behavior of the symmetrically opposite masses $\tilde M_1^2$ ($>0$) and $\tilde M_2^2$ ($<0$) showing symmetry nonrestoration at high temperature. 

It is important to point out here that:\\
{\bf (1)} typical values for $m_1^2$ and $m_2^2$ do not alter the results and conclusions since the high temperature limit is defined to be $T >> m_{1,2}^2$;\\
{\bf (2)} the model is renormalizable for any value of $a=\pm \frac{\tilde M_1^2}{\tilde M_2^2}=\pm \frac{\left[ \lambda_1 \frac{N_1 + 2}{3} + \lambda N_2 \right]}{\left[ \lambda_2 \frac{N_2 + 2}{3} + \lambda N_1 \right]}$, with 
the couplings satisfying the inequalities (\ref{eq4.1}), to preserve the 
boundness condition. In this case, the coefficients of the mass counterterms would read

\begin{eqnarray}
\label{14}
B_{1,n}= \left[ \frac{\lambda_1}{2(4\pi)^2} \left(\frac{N_1+2}{3} \right) 
 + \frac{\lambda}{2(4\pi)^2} N_2 ~ \frac{1}{a}\right]
\frac{1}{\tilde \epsilon}, \\
\nonumber 
B_{2,n}= \left[ \frac{\lambda_2}{2(4\pi)^2} \left(\frac{N_2+2}{3} \right) 
 + \frac{\lambda}{2(4\pi)^2} N_1 ~ a \right]
\frac{1}{\tilde \epsilon},
\end{eqnarray}
as shown in the Appendix.

\section{Conclusions}
\label{conc} 

In this paper we have obtained the dressed masses of the neutral multifield $O(N_1)XO(N_2)$ model at finite temperature up to one-loop order. We have used MSCR to take into account higher loops contributions and to study in what circumstances the symmetry (non)restoration can occur in a renormalized resummed theory. We have found that in the case of a multifield theory there is the necessity of the masses to obey a combination that relates them. Fortunately the $T^2$ mass behavior is a natural result of ordinary field theory at high temperature which guarantees this constraint. 

We have shown that renormalizable resummed multifield theories which exhibit the particular cases of equal masses (symmetry restoration) or symmetrically opposite masses behavior (symmetry nonrestoration) at high temperature can be constructed, depending on the physical values given to couplings which govern the strength of the interactions. In the future we want to use MSCR beyond the leading order in the perturbative expansion in order to verify the existence of other channels of ISB [like $O(N_1-1) X O(N_2-1)$], as found in \cite{Marcus1}.

\section*{Acknowledgments}
The author thanks Borut Bajc and Marcus B. Pinto for helpful comments and useful discussions about this subject and for a critical reading of the manuscript.  

\appendix
\section{Determination of the coefficients of the mass counterterms}

With MSCR the coefficients of the temperature dependent mass counterterms 
are determined as follows

\begin{eqnarray}
\label{a1}
M_{i,n}^2=(B_{i,n}+1)M_{i,n-1} ^2 + \Pi_i(M_{i,n-1}^2) \to \\
\nonumber
(B_{i,n}+1)m_i^2 + (\tilde B_{i,n}+1)\Pi_i^{Ren}(M_{i,n-2}^2)+ \Pi_i(M_{i,n-1}^2) = \\
\nonumber
(B_{i,n}+1)m_i^2 + (\tilde B_{i,n}+1)\Pi_i^{Ren}(M_{i,n-2}^2)
+c_i[ \overbrace {m_1^2+\Pi_1^{Ren} (M_{i,n-2}^2)}^{M_{1,n-1}^2} ] \frac{1}{\tilde \epsilon}\\
\nonumber
+d_i [ \overbrace {m_2^2+\Pi_2^{Ren}(M_{i,n-2}^2)}^{M_{2,n-1}^2} ] \frac{1}{\tilde \epsilon}
+ Finite_i + \Pi_i^{\beta}(M_{i,n-1}^2),
\end{eqnarray}
where $c_1=-\frac{\lambda_1}{2(4\pi)^2} \left(\frac{N_1+2}{3} \right)$ and 
$d_1=-\frac{\lambda}{2(4\pi)^2} N_2$ for $M_{1,n}^2$ and $c_2=-\frac{\lambda}{2(4\pi)^2} N_1$ and $d_2=-\frac{\lambda_2}{2(4\pi)^2} \left(\frac{N_2+2}{3} \right)$ for $M_{2,n}^2$.
As can be seen in the equation above, all divergences can be absorbed, 
provided $ M_{1,n-1}^2 (T) = aM_{2,n-1}^2 (T)+{\cal F}$. As explained 
earlier, for simplicity ${\cal F}$ is taken as zero. This implies

\begin{equation}
\label{a2}
m_2^2+\Pi_2^{Ren}(M_{1,2;n-2}^2)=\frac{1}{a} m_1^2 + 
\frac{1}{a} \Pi_1^{Ren}(M_{1,2;n-2}^2),
\end{equation}
which allows one to get the coefficients of the mass counterterms which read

\begin{eqnarray}
\label{a3}
B_{1,n}= -\left( c_1 + d_1 ~ \frac{1}{a} \right) ~ 
\frac{1}{\tilde \epsilon} ,\\
\nonumber
\tilde B_{1,n} = B_{1,n} - 1,~\forall~ n > 1,
\end{eqnarray}
and
\begin{eqnarray}
\label{a4}
B_{2,n}= -\left( a~c_2 + d_2 \right) ~ 
\frac{1}{\tilde \epsilon} ,\\
\nonumber
\tilde B_{2,n} = B_{2,n} - 1,~\forall~ n > 1.
\end{eqnarray}

It is worth noting here that at zero temperature or in the naive finite temperature calculations, there is no necessity of this constraint between the resummed thermal masses since the masses running in the loops are constants. This constraint is an effect of the resummation approach which consistently forces the masses running in the loops to be temperature dependent to fight against the breakdown of perturbative expansion. The renormalization of the resummed thermal masses at the end results in that no new counterterms are required and the Feynman rules are unchanged.



\newpage

Figure Caption

FIG. 1: The behavior of $\tilde M_i^2$. The solid line is $\tilde M_1^2$ and the dashed line is $\tilde M_2^2=-\tilde M_1^2$. This result was obtained for realistic values of the parameters of the model.

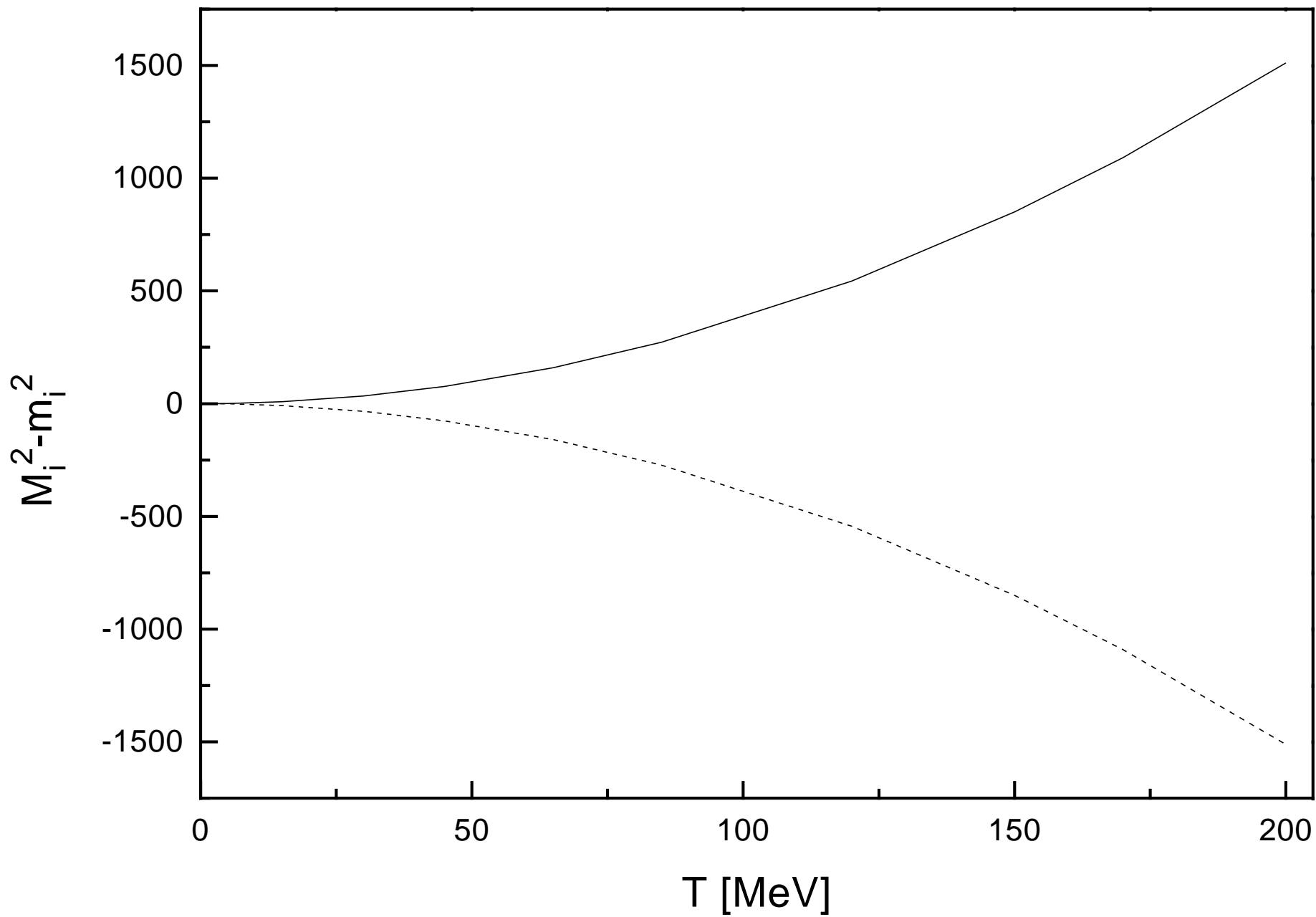
\end{document}